\documentclass{aa}
\usepackage{times,pifont}
\usepackage{graphicx}

\begin{document}

\thesaurus{03 % Extragalactic Astronomy
  (08.19.04; %supernovae: general 
  08.19.5 \object{SN\,1994D}; %supernovae: individual
  11.09.1 \object{NGC\,4526}; %galaxies: individual
  10.07.2; %globular clusters: general
  12.04.3) %distance scale
}

\title{\object{SN\,1994D} in \object{NGC\,4526}: a normally bright type
  Ia supernova}

\author{Georg Drenkhahn\inst{1,2} \and Tom Richtler\inst{1}}

\institute{Sternwarte der Universit\"at Bonn, Auf dem H\"ugel 71,
  53121 Bonn, Germany (richtler@astro.uni-bonn.de) \and
  Max-Planck-Institut f\"ur Astrophysik, Postfach 1523, 85740 Garching
  bei M\"unchen, Germany (georg@mpa-garching.mpg.de)}

\date{Received 17 March 1999 / Accepted 19 July 1999}

\maketitle

\begin{abstract}
  
  \object{SN\,1994D} of type Ia has been suspected not to fit into the
  relation between decline rate, colour, and brightness. However, an
  individual distance of its host galaxy, \object{NGC\,4526}, other
  than that of the Virgo cluster, has not yet been published. We
  determined the distance by the method of globular cluster luminosity
  functions on the basis of HST archive data.  A maximum-likelihood
  fit returns apparent turn-over magnitudes of $23.16\pm0.16$\,mag in
  $V$ and $21.96\pm0.09$\,mag in $I$. The corresponding distance
  modulus is $30.4\pm0.3$\,mag, where the error reflects our
  estimation of the absolute distance scale. The absolute magnitudes
  (not corrected for decline rate and colour) are
  $-18.67\pm0.30$\,mag, $-18.62\pm0.30$\,mag, and $-18.40\pm0.30$\,mag
  for $B$, $V$, and $I$, respectively. The corrected magnitudes are
  $-18.69\pm0.31$\,mag, $-18.69\pm0.31$\,mag, and
  $-18.44\pm0.31$\,mag.
  
  Compared with other supernovae with reliably determined distances,
  \object{SN\,1994D} fits within the errors. It is therefore not a
  counter-example against a uniform decline-rate--colour--brightness
  relation.
  
\end{abstract}

\section{Introduction}

Supernovae (SNe) of type Ia are the best standard candles in the
universe. Hamuy et~al. (\cite{hamuy:96}) showed how remarkably small
the brightness dispersion among SNe Ia is, for a sample of high
observational quality. Moreover, decline rate and colour at maximum
phase can be used to introduce corrections and to squeeze the
dispersion even below $0.15$\,mag in $B$, $V$, and $I$ (Tripp
\cite{tripp:98}; Richtler \& Drenkhahn \cite{richtler:99a}).  But a
few SNe have been observed which apparently do not fit with the
Cal\'an/Tololo sample of Hamuy et~al.; the prominent cases being
\object{SN\,1991bg} in \object{NGC\,4347}, \object{SN\,1991T} in
\object{NGC\,4527}, and \object{SN\,1994D} in \object{NGC\,4526},
which are too bright/faint/red etc.  However, checking the literature,
one finds that the claim for anomaly often relies on indirect
arguments, i.e.  without a reliable knowledge of the distance of the
host galaxy.  For example, \object{SN\,1994D} was labelled
``anomalous'' by Richmond et~al.  (\cite{richmond:95}), based on a
comparison with \object{SN\,1989B} in \object{M66} and
\object{SN\,1980N} in \object{NGC\,1316}, which made
\object{SN\,1994D} appear too bright by $0.5$\,mag. A disadvantages is
that \object{M66} only has a Tully-Fisher distance, for which it is
known that large residuals to the mean relation exist for individual
galaxies.  Moreover, this method suffers from uncertain extinction,
which, together with the dispersion in the Tully-Fisher relation,
results in an error in the absolute magnitude of the order $0.25$\,mag
(Jacoby et~al.  \cite{jacoby:92}).  On the other hand, the distance
modulus of \object{NGC\,1316} was adopted as $31.02$\,mag while
globular clusters (GCs), Cepheids, and surface brightness fluctuation
(SBF) measurements (Kohle et~al.  \cite{kohle:96}, Della~Valle et~al.
\cite{dellavalle:98}, Madore et~al. \cite{madore:98}, Jensen et~al.
\cite{jensen:98}) now agree on a value of $\sim 31.35$\,mag for the
Fornax Cluster distance modulus.  Due to the cluster's compactness we
identify the distance to \object{NGC\,1316} with the cluster distance.
An up-to-date discussion of the Fornax Cluster distance can be found
in Richtler et~al.  (\cite{richtler:99b}).

Sandage \& Tammann (\cite{sandage:95}) also found \object{SN\,1994D}
too bright by 0.25\,mag. They used the fact that \object{NGC\,4526} is
a member of the Virgo cluster and based their comparison on other
Virgo SNe. While the difference may be marginal, any correction for
the decline rate would increase the discrepancy, hence rendering the
decline-rate--luminosity relation not generally valid.  However, no
individual distance for \object{NGC\,4526} has yet been published.
The distance value used by Richmond (\cite{richmond:95}) and Patat
(\cite{patat:96}) is an unpublished SBF distance modulus of
$30.68\pm0.13$\,mag from Tonry.  In this paper, we shall give a
distance for \object{NGC\,4526}, derived by the method of globular
cluster luminosity functions (GCLFs) (e.g.  Whitmore
\cite{whitmore:97}), and we want to demonstrate that the brightness of
\object{SN\,1994D} fits reasonably well to the Fornax SNe.
        
\section{Data and reduction}

\subsection{Data}

Our data consist of HST-WFPC2 images and have been taken from the HST
archive of the European Coordinating facility at ESO. Table
\ref{tab:data} lists the exposure times and filters associated with
the data.  The date of the observation was May~8th, 1994 (Rubin
\cite{rubin:94}).  Pipe-line calibration was done with the most recent
calibration parameters according to Biretta et~al.
(\cite{biretta:96}). All frames were taken with the same pointing and
the photometry was performed on averaged frames with an effective
exposure time of 520\,s each.

Hot pixels were removed with the \textsc{IRAF} task \textsc{WARMPIX}.
The limited Charge Transfer Efficiency was corrected for with the
prescriptions given by Whitmore \& Heyer (\cite{whitmore_heyer:97}).
Cosmics were identified with the \textsc{IRAF} task \textsc{CRREJ} and
eventually the frames were combined with a weighted average.

We modelled the galaxy light with the \textsc{IRAF} task
\textsc{IMSURFIT} and achieved a very satisfying subtraction of the
galaxy light with no visible residuals.  After this, the photometric
effects of the image distortion were corrected for by a multiplication
with a correcting image (Holtzman et~al. \cite{holtzman:95}).

\begin{table}
  \caption{The names of the respective HST data set, the filter, 
    and the exposure time in seconds.}
  \label{tab:data}
  \begin{tabular}{ccr} 
    Name      & Filter & $t_\mathrm{exp}$ \\\hline
    u2dt0501t & F555W  & 60  \\
    u2dt0502t & F555W  & 230 \\
    u2dt0503t & F555W  & 230 \\
    u2dt0504t & F814W  & 60  \\
    u2dt0505t & F814W  & 230 \\
    u2dt0506t & F814W  & 230 \\\hline
  \end{tabular}
\end{table}

\subsection{Object search and photometry}

Due to the Poisson noise of the galaxy light, one finds a noise
gradient towards the centre of the galaxy.  The normally used finding
routine \textsc{DAOFIND} from the \textsc{DAOPHOT} package (Stetson
\cite{stetson:87}) does not account for a position-dependent
background noise.  Using \textsc{DAOFIND} with a mean noise level
results in many spurious detections near the galaxy centre and missed
objects in regions with lower noise background. Therefore we used the
finding routine of the \textsc{SExtractor} program (Bertin \&
Arnouts \cite{bertin:96}). The lower threshold was adjusted in order
to include also the strongest noise peaks with a number of connected
pixels of $3$ and a threshold of $1.4\cdot\sigma$.  Our initial object
list consists of 304 found and matched objects in the three wide-field
frames of the F555W and F814W images.  The planetary camera data were
not used because they showed only very few globular cluster-like
objects.

For these objects aperture photometry (with \textsc{DAOPHOT}) has been
obtained through an aperture of $0\farcs 5$. These instrumental
magnitudes were finally corrected for total flux and transformed into
standard Johnson magnitudes with the formulae given by Holtzman
et~al. (\cite{holtzman:95}).

\section{The luminosity function} 

\subsection{The selection of globular cluster candidates by 
  elongation, size and colour}

The selection of cluster candidates in HST images benefits greatly
from the enhanced resolution. While in ground-based imaging, cluster
candidates often cannot be distinguished from background galaxies,
this problem is less severe in HST images. Galaxies that are so
distant that their angular diameter is comparable to cluster
candidates in \object{NGC\,4526}, normally have quite red colours.
Remaining candidates are foreground stars and faint background
galaxies, where the redshift balances an intrinsic blue colour. Our
strategy, described in more detail below, is the following: we first
select our sample with respect to the elongation as given by the
\textsc{SExtractor} program, then the angular size measured as the
difference between two apertures, and finally the colour.

\subsubsection{Elongation}

The elongation is taken as the ratio $E=A/B$, where $A$ and $B$ are
the major and the minor axes as given by \textsc{SExtractor}.
However, the undersampling can cause a spurious elongation if a given
object is centred between 2 pixels. We therefore chose $E<1.9$ in both
filters to exclude only the definitely elongated objects.

\subsubsection{Size}

The \textsc{SExtractor} stellarity index is not very suitable for
undersampled images because the neural network of \textsc{SExtractor}
is trained with normally sampled objects (Bertin \& Arnouts
\cite{bertin:96}). We use the magnitude difference in two apertures as
a size measure (Holtzman et~al. \cite{holtzman:96}).  These apertures
should be compatible with the expected size of a GC in
\object{NGC\,4526}. A typical half-light radius of a GC is of the
order of a few pc. Thus the larger clusters are marginally resolvable,
assuming a distance of 13\,Mpc to \object{NGC\,4526}, which
corresponds to 6\,pc per pixel.  The parameter
$m_{12}:=m_1-m_2-(m_1-m_2)_*$ is defined as the magnitude difference
between one and two pixel apertures $m_1-m_2$ minus the appropriate
value for a star-like object $(m_1-m_2)_*$. The latter subtraction
accounts for the position-dependent PSF of the WFPC2.

\begin{figure}
  \includegraphics[width=\hsize]{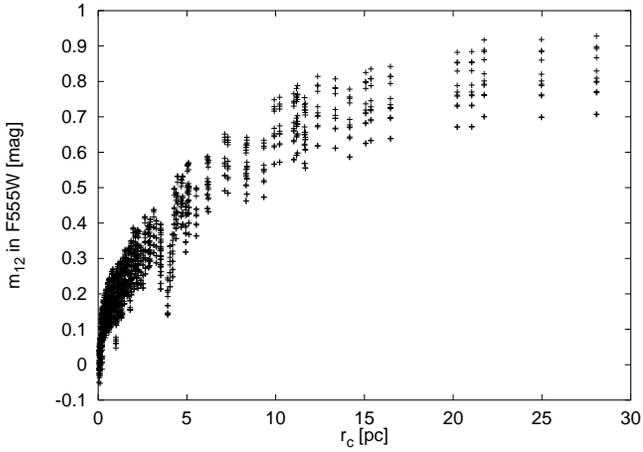}
  \caption{The measured $m_{12}$ values of the simulated Galactic GCs as a
    function of the core radius $r_c$.
% The core radius $r_c$ of the simulated Galactic GCs 
% as a function of the measured $m_{12}$ values.
    Every GC is represented by 12 crosses according to the 3 WF
    cameras and 4 different subpixel positions.}
  \label{fig:rcm12}
\end{figure}

To find a reasonable size limit, we simulated cluster candidates by
projecting the data for Galactic GCs at a distance of 13\,Mpc, based
on the McMaster catalogue of Galactic globular clusters (Harris
\cite{harris:96}). We generated oversampled King-profiles (King
\cite{king:62}) with the geometric parameters from this catalogue and
convolved them with an oversampled PSF of the WFPC2. The resulting
images were rebinned at four different subpixel positions giving WFPC2
images of the Galactic globular cluster system (GCS) as it would be
observed from a distance of 13\,Mpc. Fig.~\ref{fig:rcm12} shows the
measured $m_{12}$ values as functions of the core radii $r_c$. It
turns out that $m_{12}$ is a sufficiently good quantity to measure
$r_c$.
Applying a selection criterion of $m_{12}<0.45$ in both filters
excludes extended GCs with $r_c>5$\,pc. This reduces the size of the
GC sample but does not introduce a bias because $r_c$ and $M_V$ are
not correlated.  The $m_{12}$ magnitude is not suitable to place a
lower size limit because the smaller GCs are unresolved and cannot be
distinguished from a stellar-like image.

\begin{figure}
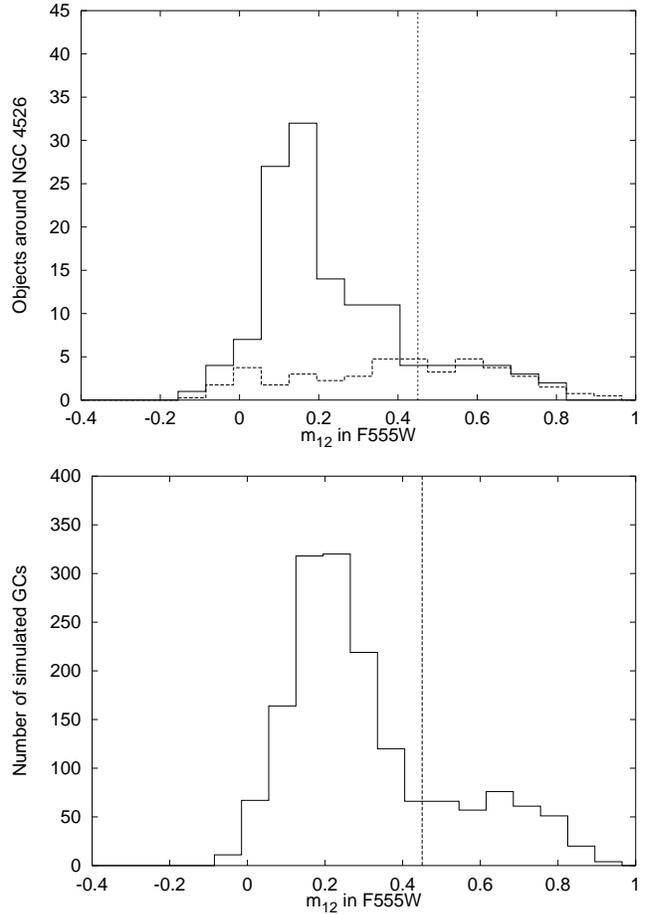

  \includegraphics[width=\hsize]{8653.f2}\\
  \includegraphics[width=\hsize]{8653.f3}
  \caption{These histograms show the distribution of $m_{12}$ for 
    the detected sources (upper panel) and the simulated clusters
    (lower panel). The dotted histogram shows the expected background
    objects (=galaxies) derived from the Medium Deep Survey frames.
    Objects with values $m_{12}>0.45$ are resolved and mostly
    background objects. The distribution for F814W is essentially the
    same.}
  \label{fig:m12}  
\end{figure}

Fig.~\ref{fig:m12} compares the distribution of $m_{12}$ of the sample
of our detected sources with that of the simulated clusters for both
filters. The dotted histogram represents the background sources (see
Sec.~\ref{sec:bg}). The objects with $m_{12}>0.45$ can be almost
entirely explained as resolved background objects.

\subsubsection{Colours}
\begin{figure}
  \includegraphics[width=\hsize]{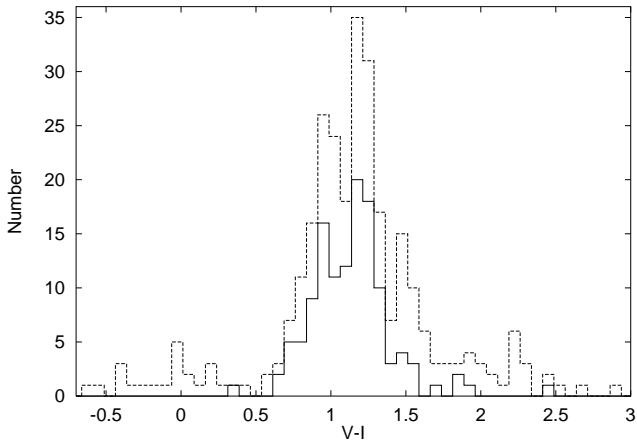}
  \caption{This colour histogram shows the shape- and size-selected 
    sample of objects. The dotted line denotes the distribution of the
    complete sample. It can be seen that the shape and size
    distribution can already discriminate against red objects which
    are presumably distant galaxies. For the final sample, we chose
    the colour interval $0.8<V-I<1.4$.}
  \label{fig:colours} 
\end{figure}

The range of existing colours among Galactic GCs is not a very good
guide in our case: large error bars and a possibly different
metallicity distribution may cause a different shape of the colour
histogram.  However, as Fig.~\ref{fig:colours} shows, a selection with
regard to shape and size selects also according to colours, i.e. the
redder objects, which are presumably redshifted background galaxies,
can be effectively removed.

The colour histogram of the remaining objects shows some similarity to
other GCSs with regard to the apparent ``bimodality'', for example in
\object{M87} (Whitmore et~al.  \cite{whitmore:95}). One may suspect
that there is a dichotomy in the population of GCs in the sense that
the blue (=metal poor) objects belong to a halo population and the red
ones to the bulge.  Unfortunately, the number of objects found does
not allow a deeper investigation. The mean colour is $\langle
V-I\rangle=1.1$.

We also exclude very faint objects with large photometric errors
$\Delta m>0.3$ in at least one of the two filters. They hardly
contribute to the determination of the turn-over magnitude (TOM)
because of their large uncertainty.

\subsection{Completeness}

The completeness $I(m) \cdot \mathrm{d}m$ is the probability of
finding an object in the magnitude interval $[m,m+\mathrm{d}m]$. On
frames with a homogeneous background, it depends mainly on the
brightness of an object. In our case, where the object search was done
on a frame with a subtracted galaxy, the noise of the background
$\rho$ varies strongly, and must be treated as a second parameter in
$I(m;\rho)$.

We evaluated the completeness in our $V$, $I$ sample by inserting
artificial objects in the $V$-frame and we used the same coordinates
for inserting objects in the $I$-frame, after having fixed the colour
for all objects to the value of $V-I=1.1$.  In total we used 19000
artificial ``stars'' in the magnitude range $23<V<25.5$ and evaluated
the completeness in bins of 0.1\,mag. The noise as the second
parameter is taken to be the standard deviation $\rho$ of pixel values
in the annulus used by the aperture photometry. In practice, we used
three completeness functions for three different $\rho$-values
$1.75$\,DN, $2.05$\,DN, and $3.35$\,DN, which represent the intervals
$[1.6,1.9]$, $[1.9,2.2]$, and $[2.2,4.5]$, respectively.

\begin{figure}
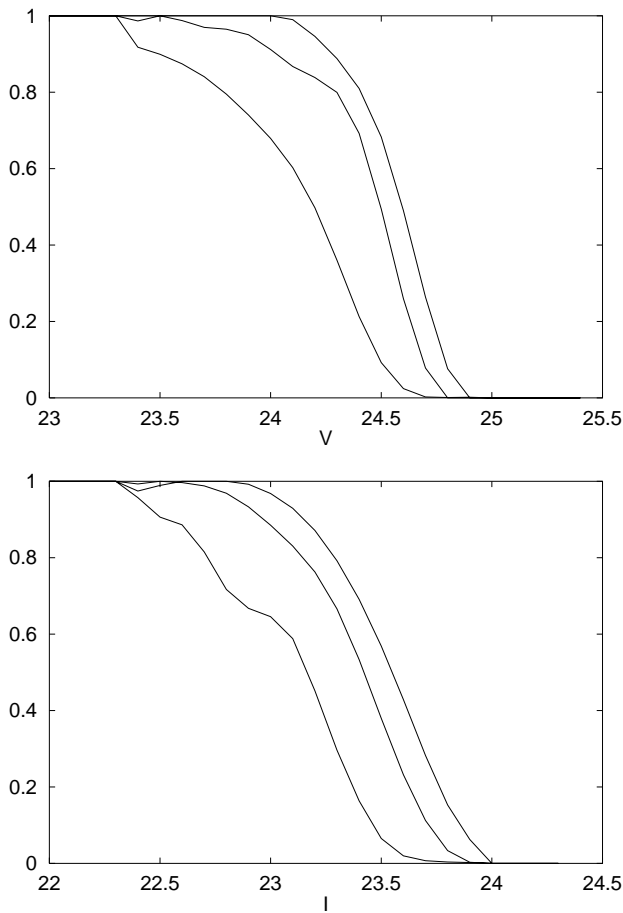

  \includegraphics[width=\hsize]{8653.f5}\\
  \includegraphics[width=\hsize]{8653.f6}
  \caption{The completeness for $V$ and $I$ for three different noise 
    intervals, which are characterised by the mean values
    $\rho=1.75$\,DN, $2.05$\,DN and $3.35$\,DN (from right to left) in
    the $V$ filter.}
  \label{fig:comp} 
\end{figure}

Fig.~\ref{fig:comp} shows the completeness for the three noise
intervals and for the two filters. The distinct difference between the
three curves demonstrates that $\rho$ is indeed a reasonable
parametrisation.

\subsection{Foreground and background sources}
\label{sec:bg}

The field of our HST images is too small for a proper analysis of the
contaminating background sources that survived all selection
processes.

Therefore, we used the WFPC2 Medium Deep Survey (MDS) (Ratnatunga
et~al. \cite{ratnatunga:99}) to obtain an estimation of the
background.  Suitable fields should be possibly nearby, in particular
regarding the Galactic latitude, and should be deep with several
single exposures.  We found the exposures which are listed in
Table~\ref{tab:MDS} together with some information. The frames are
deep enough (in F606W and F814W) that the source finding is complete
in the relevant magnitude interval. $V$ and $I$ magnitudes are
calculated according to the prescriptions of Holtzman et~al.
(\cite{holtzman:95}).

\begin{table}
  \caption{The exposures from the Medium Sky Survey which 
    have been used for the background evaluation. Given are the data
    set names, total exposure times in seconds, and the number 
    of individual frames, which entered the averaging process.}
  \label{tab:MDS}
  \begin{tabular}{crrcc} 
    \hline
     & $t_\mathrm{exp}$ [s] & $t_\mathrm{exp}$ [s] & \# & \#\\
    Name  & in F606W & in F814W & in F606W & in F814W\\\hline 
    u30h3 & 1200  & 1266  & 6  & 8 \\
    uy400 & 866   & 1000  & 6  & 6 \\
    uy401 & 600   & 700   & 4  & 8 \\
    uyx10 & 800   & 700   & 3  & 3 \\ 
    uzx01 & 696   & 1550  & 5  & 4 \\\hline
  \end{tabular}
\end{table}

The selection criteria discussed above are applied in exactly the same
way as in the case of the \object{NGC\,4526} frames. It turns out that
most background objects are selected out. The Poisson scatter of
background objects from frame to frame is in the expected statistical
range. Thus any clustering in the background galaxy number density
does not influence the background LF. So we can assume that the
background LF derived from the MDS frames gives also a good estimate
for the background LF near \object{NGC\,4526.}

As can be assessed also from Fig.~\ref{fig:m12}, the fraction of
background objects in the selected samples is only 7\%. The resulting
luminosity function $b(m)$ enters the maximum-likelihood fitting
process as described in Sec.~\ref{sec:pdf}.

\subsection{The turn-over magnitude}

\subsubsection{Remarks on the maximum-likelihood fit and the TOM 
  of \object{NGC\,4526}}
\label{sec:pdf}

For the representation of the GCLF, we choose the $t_5$-function,
which has been shown to be the best analytical representation of the
Galactic GCLF and that of \object{M31} (Secker \cite{secker:92}). It
reads
\begin{equation}
  \label{eq:t5}
  t_5(m;m^0,\sigma_t) = 
  \frac{8}{3\sqrt5\pi\sigma_t}
  \left(1+\frac{(m-m^0)^2}{5\sigma_t^2}\right)^{-3}.
\end{equation}
The maximum-likelihood fitting method evaluates the most probable
parameters for a given distribution $\phi$ by maximising the
likelihood
\begin{equation}
  \label{eq:L}
   L(m^0,\sigma_t) = \prod_i \phi(m_i;m^0,\sigma_t).
\end{equation}
For the theoretical background of this method see Bevington \&
Robinson (\cite{bevington:92}), Caso et~al. (\cite{caso:98}). Secker
(\cite{secker:92}) and Secker \& Harris (\cite{secker:93}) already
applied it to the fitting of GCLFs.

What is to be fitted is the $t_5$-function modified by the different
influencing factors, i.e. the completeness, the background, and the
photometric errors. We used an extended version of the distribution of
Secker (\cite{secker:92}) that accounts for a position-dependent
background noise.
\begin{equation}
  \label{eq:phi}
  \begin{array}{l}
    \phi(m;\rho,m^0,\sigma_t)=\\
    \quad K\cdot\kappa(m;\rho,m^0,\sigma_t)
    +(1-K)\cdot\beta(m;\rho)
  \end{array}
\end{equation}
with $\kappa$ being the normalised distribution of the GCs
\begin{equation}
  \label{eq:kappa}
  \begin{array}{l}
    \kappa(m;\rho,m^0,\sigma_t)=\\
    \quad I(m;\rho)\cdot\left(\int
      t_5(m';m^0,\sigma_t)\cdot
      \varepsilon(m;m',\rho)\,\mathrm{d}m' \right)
  \end{array}
\end{equation}
and $\beta$ being the normalised distribution of the background
objects
\begin{equation}
  \label{eq:beta}
  \beta(m;\rho)=I(m;\rho)\cdot b(m).
\end{equation}
Special care must be taken when $\kappa$ and $\beta$ are normalised
and mixed through the mixing value $K$, because this depends on the
completeness and therefore on $\rho$.
 
$\varepsilon$ is a Gaussian distribution accounting for the
photometric error
\begin{equation}
  \label{eq:eps}
  \varepsilon(m;m',\rho)=\frac{1}{\sqrt{2\pi}\cdot\sigma(m',\rho)}
  \cdot\exp\left(\frac{(m-m')^2}{2\cdot\sigma(m',\rho)}\right)
\end{equation}
with a magnitude- and noise-dependent width
\begin{equation}
  \label{eq:sig}
  \sigma(m,\rho)=\sqrt{10^{0.4\cdot(m-a)}
    +\rho^2\cdot 10^{0.8\cdot(m-b)}}
\end{equation}
as used by \textsc{DAOPHOT} (Stetson \cite{stetson:87}). $a$ and $b$
depend on the photometry parameters but are more easily determined by
just fitting the photometry data to (\ref{eq:sig}).

The necessity for this more complicated approach is most clearly seen
in the noise-dependent completeness functions in Fig.~\ref{fig:comp}.
One could also perform the analysis for objects in a small range of
background noise but this would further reduce our already small GC
sample.

\begin{figure}
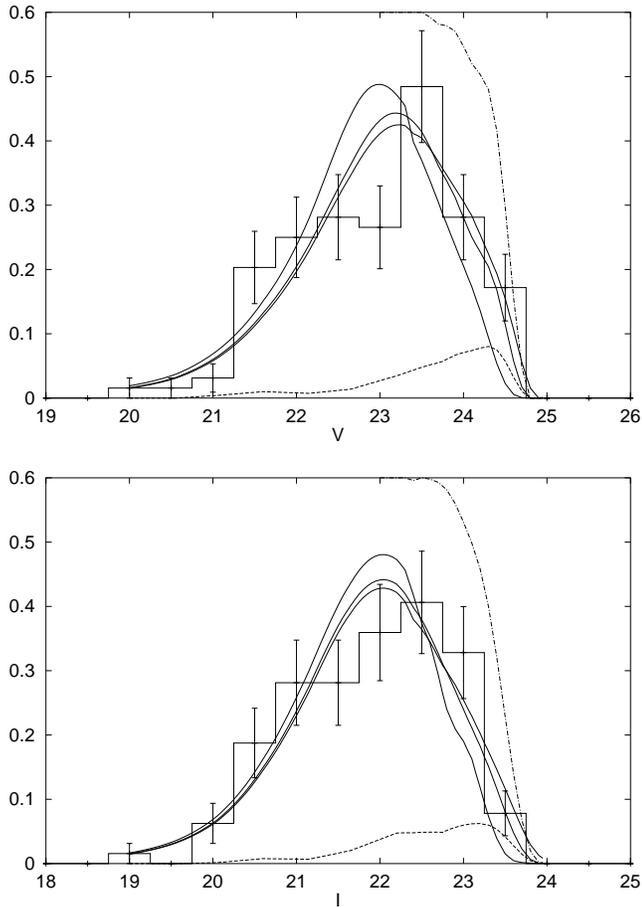

  \includegraphics[width=\hsize]{8653.f7}\\
  \includegraphics[width=\hsize]{8653.f8}
  \caption{The result of the maximum-likelihood fit for the $V$ and
    $I$ filter (solid line). Histogram: all selected objects
    independent of their $\rho$ values.  Solid lines: fitted GCLFs
    $\phi(m;1.75$\,DN$)$, $\phi(m;2.00$\,DN$)$, and
    $\phi(m;3.25$\,DN$)$.  Dotted-dashed line: down-scaled
    completeness $0.6\cdot I(m;2$\,DN$)$.  Dashed line: background
    contribution $\beta(m;2$\,DN$)$. The ordinate is the dimensionless
    probability per magnitude for $\phi$, $\beta$ and the histogram
    while for $I$ it just denotes the probability. As required by any
    probability density function the area under the fitted
    distribution function $\phi(m;\rho)$ and the histogram is unity
    for all values of $\rho$.}
  \label{fig:TOM}
\end{figure}

Fig.~\ref{fig:TOM} shows two-dimensional slices through the
three-dimensionally fitted luminosity function at a noise level of
$\rho=1.75$, $2.00$, and $3.25$\,DN, respectively, in both filters.
The background noise of $\rho=2.00$\,DN corresponds approximately to
the mode of the background noise distribution and is therefore most
representative.  The figure also displays the low background
contribution and a down-scaled completeness function for
$\rho=2.00$\,DN.  The histogram includes all objects independent of
their $\rho$-values. It therefore does not correspond to a single
fitted function but is rather a superposition of many functions
belonging to a variety of $\rho$-values.  Though the histogram itself
is not used in the fitting it shows that the fitted distribution
models the data reasonably well. From simply looking at the histogram
one could conclude that the TOM lies at a slightly fainter magnitude.
But one has to keep in mind the large statistical scatter within the
single bins and that this picture changes with different bin positions
and widths. Note also the differences between the maxima of the
functions $\phi(m;1.75$\,DN$)$, $\phi(m;2.00$\,DN$)$, and
$\phi(m;3.25$\,DN$)$ in $V$ caused by the different completeness
limits. The completeness function in our data decreases at the same
magnitude value as the declining part of the intrinsic LF. It is
therefore crucial to determine the completeness function most
precisely and any error in this part of the calculation introduces a
significant systematic error in the TOM.  This problem is only
resolved if the observation is complete down to at least one magnitude
beyond the TOM.

\begin{figure}
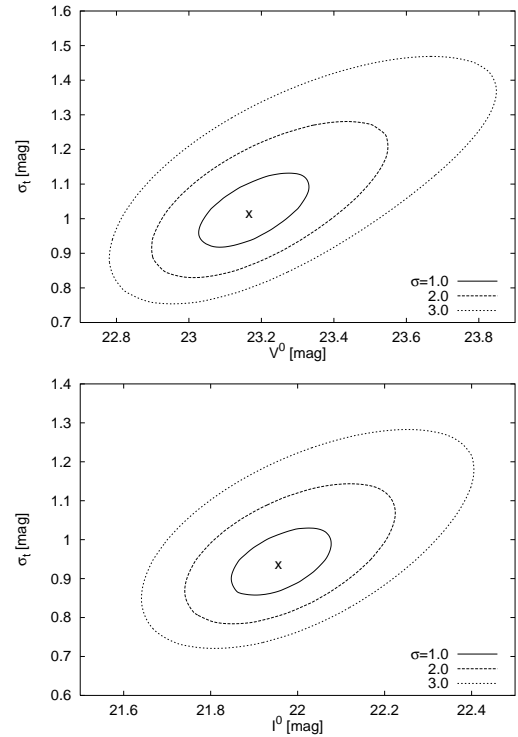

  \begin{center}
    \includegraphics[width=0.8\hsize]{8653.f9}\\
    \includegraphics[width=0.8\hsize]{8653.f10}
  \end{center}
  \caption{1, 2, and $3\sigma$ contours for the fitted parameters 
    $m^0$ and $\sigma_t$ in $V$ and $I$.}
  \label{fig:cont}
\end{figure}

Fig.~\ref{fig:cont} shows the likelihood contours for the fitted
parameters and Table~\ref{tab:TOM} gives the resulting values.
The width $\sigma_t$ is $0.09$\,mag larger than $\sigma_t$ for the
Galactic system (Table~\ref{tab:galac}) but still small compared to
other early-type galaxies (Whitmore \cite{whitmore:97} use
$\sigma_t=0.78\cdot\sigma_\mathrm{g}$ to convert the Gaussian widths).
 
\begin{table}
  \caption{The fitted turn-over magnitudes and $t_5$ dispersions for 
    the globular cluster system of \object{NGC\,4526}.}
  \label{tab:TOM}
  \begin{tabular}{ccc} 
    Filter & $m_0$          & $\sigma_t$ \\\hline
    $V$    & $23.16\pm0.16$ & $1.01\pm0.12$ \\
    $I$    & $21.96\pm0.09$ & $0.94\pm0.09$ \\\hline
  \end{tabular}
\end{table}

\subsubsection{The TOM of the Galactic system}  

Although the Galactic TOM has been discussed by many authors, it is
not a parameter that can be simply looked up in the literature. The
exact value depends on the method of fitting, on the selection of GCs
that enter the fitting process, on the distance scale of GCs, and on
the exact application of a reddening law. Even if our values do not
differ much from previous evaluations, we discuss them for numerical
consistency.

We take the LMC distance as the fundamental distance for calibrating
the zero-point in the relation between metallicity and horizontal
branch/RR\,Lyrae brightness.  The third fundamental distance
determination beside trigonometric parallaxes and stellar stream
parallaxes is the method of Baade-Wesselink parallaxes. It has been
applied to the LMC in its modified form known as the Barnes-Evans
method.  So far, it has been applied to Cepheids in \object{NGC\,1866}
(Gieren et~al. \cite{gieren:94}), and the most accurate LMC distance
determination to date stems from the period-luminosity relation of LMC
Cepheids by Gieren et al.  (\cite{gieren:98}). We adopt the distance
modulus from the latter work, which is $18.46\pm0.06$\,mag, and which
is in very good agreement with most other work (e.g. Tanvir
\cite{tanvir:96}).

If we adopt the apparent magnitude of RR\,Lyrae stars in the LMC from
Walker (\cite{walker:92}), $18.94\pm0.1$\,mag for a metallicity of
$[\element{Fe}/\element{H}]=-1.9$\,dex, and the metallicity dependence
from Carretta et~al. (\cite{carretta:99}), we find
\begin{equation}
  \label{eq:mvrr}
   M_V(RR) = (0.18\pm0.09)([\element{Fe}/\element{H}]+1.6)+0.53\pm0.12.
\end{equation}
This zero-point is in excellent agreement with the value of
$0.58\pm0.12$\,mag derived from HB-brightnesses of old LMC globular
clusters from Suntzeff et~al. (\cite{suntzeff:92}), if the above
metallicity dependence is used.
%%% Suntzeff: V(HB)=18.99\pm0.09 bei [Fe/H]=-1.6 und A_V/E(B-V)=3.1
It also agrees within a one-$\sigma$ limit with $0.56\pm0.12$\,mag
(Chaboyer \cite{chaboyer:99}) and $0.57\pm0.04$\,mag (Carretta et~al.
\cite{carretta:99}).
 
We calculate the absolute magnitudes of Galactic GCs from the McMaster
catalogue (Harris \cite{harris:96}), using $A_V/E(B-V)=3.1$ (Rieke \&
Lebofsky \cite{rieke:85}) and $E(V-I)/E(B-V)=1.53$ (Ardeberg \&
Virdefors \cite{ardeberg:82}) as a compromise between the extreme
values $1.6$\,mag (Rieke \& Lebofsky \cite{rieke:85}) and $1.35$\,mag
(Drukier et~al.  \cite{drukier:93}).  We exclude clusters with
$E(B-V)>1$ so that the sample embraces 120 clusters in $V$ and 93 in
$I$.

\begin{figure}
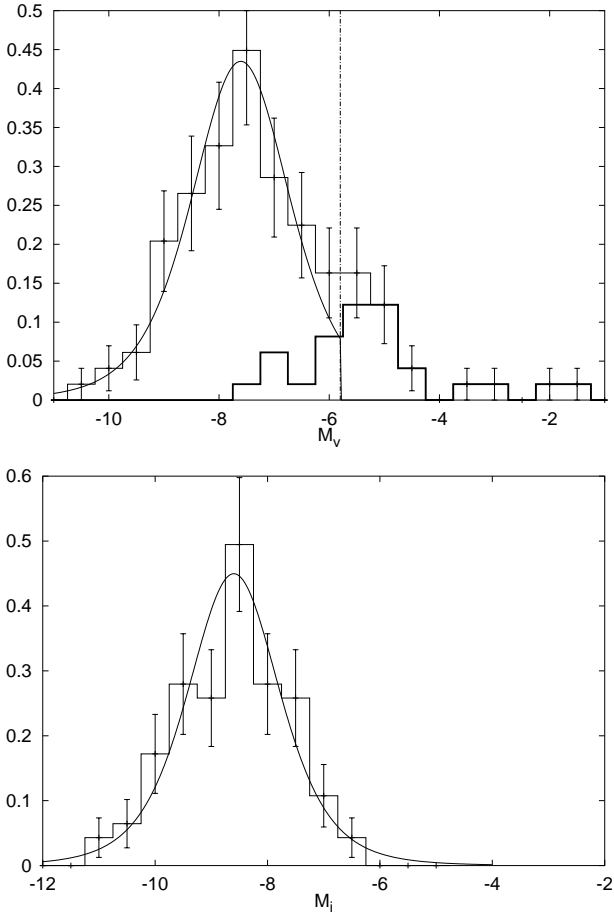

  \includegraphics[width=\hsize]{8653.f11}\\
  \includegraphics[width=\hsize]{8653.f12}
  \caption{This graph shows the normalised distribution of absolute 
    magnitudes in $V$ and $I$ for Galactic globular clusters.
    Overplotted as solid lines are the $t_5$-functions resulting from
    a maximum-likelihood fit. The bold histogram in the left panel
    shows the GCs with missing $I$-band photometry. The dotted
    vertical line denotes the cut-off magnitude at $m_c=-5.8$.}
  \label{fig:galac}
\end{figure}

Fig.~\ref{fig:galac} shows the scaled histograms ($V$ and $I$) for the
sample of Galactic globular clusters. The solid line is the
$t_5$-function resulting from the maximum-likelihood fit. It is not
the fit to the histograms, which are shown just for visualisation
purposes.

\begin{figure}
  \begin{center}
    \includegraphics[width=\hsize]{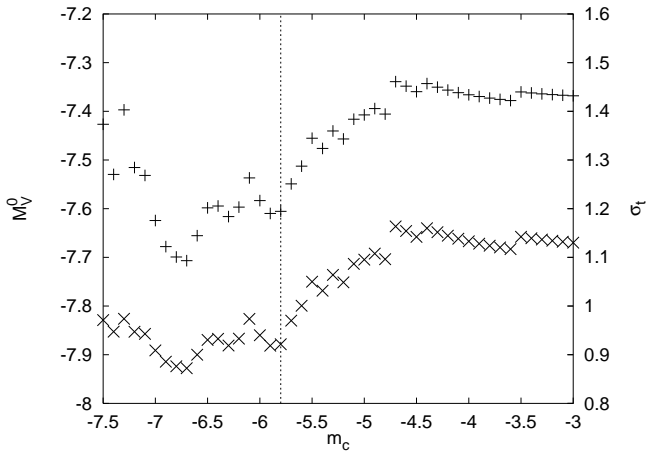}
    \caption{Dependence of the fitted $M_V^0$ (symbol \ding{53}) and
      $\sigma_t$ (symbol $+$) on the cut-off magnitude $m_c$ for the
      Galactic GCS. The chosen $m_c=-5.8$ is indicated by the dashed
      line.}
    \label{fig:grenze}
  \end{center}
\end{figure}

It is apparent from Fig.~\ref{fig:galac} that the symmetry in the $V$
distribution breaks down for clusters at the faint end of the
distribution.  It is reasonable to include in the fit only the
symmetric part up to a certain limiting magnitude. To determine an
appropriate cut-off magnitude $m_c$ we carry out the
maximum-likelihood fit for various $m_c$ and look at how the fitted
parameters evolve.  Fig.~\ref{fig:grenze} shows that the $V$-TOM
$M^0_V$ and the width $\sigma_t$ clearly depend on $m_c$. The
inclusion of the asymmetric part shifts $M_V^0$ systematically to
greater values. This trend stops at our adopted value $m_c=-5.8$\,mag
and lowering $m_c$ further just results in a random scatter of
$M_V^0$.

\begin{table}
  \caption{Fitted TOMs and $\sigma_t$-values for the Galactic GC
    system. The $I$ values are biased due to selection effects
    as discussed in the text.}
  \label{tab:galac}
  \begin{tabular}{cccc}
    Filter & $M^0$          & $\sigma_t$    & \# of GCs\\\hline 
    $V$    & $-7.61\pm0.08$ & $0.92\pm0.07$ & 120 \\
    $I$    & $-8.60\pm0.07$ & $0.85\pm0.06$ & 93  \\\hline
  \end{tabular}
\end{table}

There is no asymmetric part in the histogram of the $I$ band data in
Fig.~\ref{fig:galac}. The reason for this is a selection effect in the
$I$ photometry. Looking at the McMaster catalogue data reveals missing
$I$ data in approximately 22\% of the listed GCs. All of these
clusters with missing data belong to the fainter end of the luminosity
function in $V$. This is shown by plotting the subset of GCs without
$I$ magnitudes onto the $M_V$ histogram in Fig.~\ref{fig:galac}. This
incompleteness shifts the fitted $I$ TOM to brighter magnitudes. Thus,
a comparison between the difference of the fitted TOMs and the mean
colour of the Galactic GCs gives
\begin{equation}
  \label{eq:vishift}
  0.87\pm0.09 = \langle V-I\rangle_\mathrm{MW} 
  < M_V^0 - M_I^0 = 1.01\pm0.11.
\end{equation}
The median is equal to the mean of the $V-I$ distribution for the
Galactic GCs.  We therefore use
\begin{equation}
  \label{eq:itom}
  M_I^0=M_V^0-\langle V-I\rangle_\mathrm{MW} = -8.48\pm0.10 
\end{equation}
as an absolute TOM in the $I$ band.  Alternatively, we introduce a
cut-off magnitude in the $I$ band data and fit only the more complete
luminous part of the distribution like we did in the $V$ band.  As
expected, the fitted $I$ TOM becomes smaller with decreasing cut-off
magnitude.  If the cut-off magnitude approaches the TOM the fitted TOM
values begin to scatter greatly around approximately $-8.52$\,mag
because the sample size decreases and the fit results are less well
determined. We therefore do not rely on this approach but rather
regard it as a confirmation of the validity of (\ref{eq:itom}).

\subsection{Metallicity correction and distance}

The luminosity of a GC depends not only on its mass but also on its
metallicity.  Any difference between the metallicities of the GCS of
\object{NGC\,4526} and the Galactic GCS must be corrected to avoid a
systematic bias.

The amount of this correction has been studied by Ashman et~al.
(\cite{ashman:95}) and we adopt their values. The only metallicity
indicator that is available for us is the mean $V-I$ colour of the
GCS, for which we adopt the relation given by Couture et~al.
(\cite{couture:90}):
\begin{equation}
  \label{eq:vi}
   V-I = 0.2\cdot [\element{Fe}/\element{H}]+1.2.
\end{equation}
A mean colour of $\langle V-I\rangle = 1.1$ indicates
$[\element{Fe}/\element{H}]_\mathrm{NGC4526}=-0.5\pm0.3$ and we find a
metallicity correction of $\Delta M_V^0 = 0.28\pm0.19$ and $\Delta
M_I^0 = 0.11\pm0.05$. The large error of $\pm0.3$\,dex for the
metallicity arises from the fact that $\langle V-I\rangle$ depends
only weakly on the metallicity. This value can be estimated from
looking at the scatter in the $[\element{Fe}/\element{H}]$--$(V-I)$
plot of the Galactic GCS.

Hence the distance moduli in $V$ and $I$ are
\begin{equation}
  \label{eq:mmvi}
  % hier sollten wir \mu_V statt (m-M)_V verwenden, denn der
  % Entfernungsmodul enthaelt nicht nur 2 Komponenten hier.
  \mu_V=30.49\pm0.26
  %=(23.16\pm0.16)-(-7.61\pm0.08)-(0.28\pm0.19)
  \quad\mbox{and}\quad
  \mu_I=30.33\pm0.14.
  %=(21.96\pm0.09)-(-8.48\pm0.10)-(0.11\pm0.05)
\end{equation}
If the universality of the TOM of $\approx0.2$\,mag (Whitmore
\cite{whitmore:97}) and the uncertainty in the horizontal branch
calibration of $\approx0.1$\,mag is included as external errors we
obtain
\begin{equation}
  \label{eq:mmngc4526}
  \mu_\mathrm{NGC4526}=30.4\pm0.3\,\mbox{mag}
\end{equation}
and 
\begin{equation}
  \label{eq:Dngc4526}
  D_\mathrm{NGC4526}=12.0\pm1.6\,\mbox{Mpc}
\end{equation}
as our final value for the distance modulus and distance. 

Thus, \object{NGC\,4526} is located in the foreground of the Virgo
cluster and a comparison with apparent magnitudes of other SNe in the
Virgo cluster is misleading. This result is consistent within
$1\sigma$ with Tonry's unpublished SBF distance modulus
$30.68\pm0.13$\,mag, which was used by Richmond et~al.
(\cite{richmond:95}).

\section{Comparison with other supernovae}

\subsection{The dependence on decline rate and colour}

The publication of the Cal\'an/Tololo-sample of Ia SNe (Hamuy et~al.
\cite{hamuy:96}) confirmed in a compelling way what has been
suggested earlier, namely that the luminosity of Ia's depends on the
decline rate and also on the colour of the SN at maximum phase.

\begin{figure}
  \includegraphics[width=\hsize]{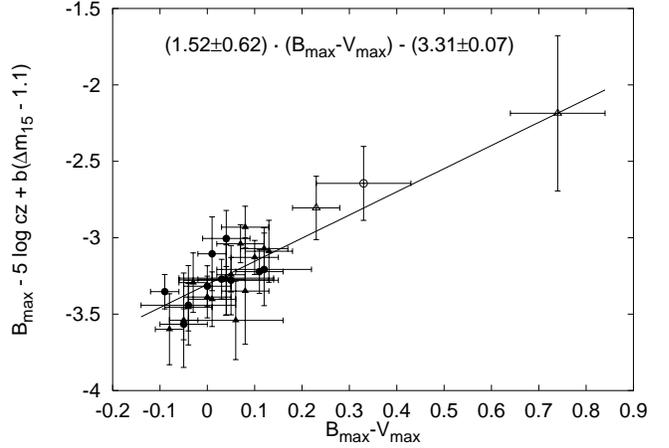}
  \caption{This plot shows the relation between 
    $B_\mathrm{max}-V_\mathrm{max}$ rate and the apparent maximum
    brightness in $B$ corrected for the decline rate. A clear
    dependence is visible. We point out that only blue SNe with
    $B_\mathrm{max}-V_\mathrm{max}<0.2$ have been included in the fit.
    The fact that the redder SNe are well represented by this
    relation is a point in favour of the validity of this relation
    even over a larger colour range.}
  \label{fig:colcor} 
\end{figure}

Here we report convenient expressions which relate in the sample of
Hamuy et~al. the decline rate $\Delta m_{15}$ and the foreground
extinction-corrected colour $B_\mathrm{max}-V_\mathrm{max}$ with the
apparent maximum magnitude (the recently published sample by Riess
et~al. (\cite{riess:98}) does not change significantly the numerical
values).  These linear relations read
\begin{equation}
  \label{eq:mcor}
  \begin{array}{l}
    m_{\mathrm{max},i} + b \cdot (\Delta m_{15} -1.1) 
    + R \cdot(B_\mathrm{max}-V_\mathrm{max})\\
    ~ = 5 \cdot\log cz + Z
  \end{array}
\end{equation}
where $i\in \{B,V,I\}$ denotes the different bands. Table
\ref{tab:coeff} lists the coefficients, which have been obtained by a
maximum-likelihood fit, together with their errors. Only SNe with
$B_\mathrm{max}-V_\mathrm{max}<0.2$ have been used for the fit.
Fig.~\ref{fig:colcor} shows the relation between
$B_\mathrm{max}-V_\mathrm{max}$ and the apparent maximum brightness
corrected for decline rate. The fact that the red SNe are also well
fitted by this relation indicates the validity of this relation for a
larger colour range.  The coefficients of the colour term also deviate
strongly from the Galactic reddening law so that one may conclude that
the red colours are indeed intrinsic and not significantly affected by
reddening.

The corrected magnitudes can now be compared with those of other SNe.
 
\begin{table}
  \caption{The table lists the coefficients according to
    (\ref{eq:mcor}) for the relation between brightness, 
    decline rate, and colour for the Cal\'an/Tololo sample.}
  \label{tab:coeff}
  \begin{flushleft}
    \begin{tabular}{cccc} 
      Band  & $Z$              & $b$            & $R$ \\\hline
      $B$   & $-3.306\pm0.063$ & $-0.48\pm0.23$ & $-1.51\pm0.62$ \\ 
      $V$   & $-3.320\pm0.058$ & $-0.52\pm0.19$ & $-0.83\pm0.57$ \\ 
      $I$   & $-3.041\pm0.060$ & $-0.40\pm0.22$ & $-0.81\pm0.59$ \\ 
      \hline
    \end{tabular}
  \end{flushleft}
\end{table}

\subsection{Comparison with other SNe}

\begin{table*}[ht]
  \caption{The observational data and distance moduli of the SNe. 
    The first three rows show the extinction-corrected apparent
    magnitudes computed with $A_B/E(B-V)=4.09$, $A_V/E(B-V)=3.09$ and
    $A_I/E(B-V)=1.49$ (Rieke \& Lebofsky \cite{rieke:85}).
    Photometric data, colour excesses and decline rates are taken from
    Richmond et~al.  (\cite{richmond:95}) for \object{SN\,1994D}, from
    Hamuy et~al. (\cite{hamuy:96}) for \object{SN\,1992A} and
    \object{SN\,1980N}, and from Hamuy et~al. (\cite{hamuy:91}) for
    \object{SN\,1981D}. The absolute magnitudes are corrected for
    decline rate and colour by (\ref{eq:mcor}).}
  \label{tab:sndata}
  \begin{tabular}{rrrrr}
    & \object{SN\,1994D}&\object{SN\,1992A}&\object{SN\,1980N}
    &\object{SN\,1981D}\\
    \hline
    $B_\mathrm{max}$ & $11.73\pm0.02$ & $12.57\pm0.03$ &
      $12.49\pm0.03$ & $12.59$\\
    $V_\mathrm{max}$ & $11.78\pm0.02$ & $12.55\pm0.03$ & 
      $12.44\pm0.03$ & $12.40$\\
    $I_\mathrm{max}$ & $12.00\pm0.05$ & $12.80\pm0.03$ & 
      $12.70\pm0.04$ & ---\\
    $\Delta m_{15}(B)$& $1.31\pm0.08$ &  $1.47\pm0.05$ &  
      $1.28\pm0.04$ & ---\\
    $B_{\mathrm{max}}-V_{\mathrm{max}}$ 
                     & $-0.05\pm0.04$ & $0.02\pm0.04$  & 
                     $0.05\pm0.04$ & $0.19$\\
    $E(B-V)$         &  $0.04\pm0.03$ & $0$            & 
      $0$ & $0$ (?)\\
    $\mu$            & $30.40\pm0.30$ & $31.35\pm0.15$ & 
      $31.35\pm0.15$ & $31.35\pm0.15$\\
    \hline
    $M_{B,\mathrm{cor}}$ & $-18.69\pm0.31$ & $-18.99\pm0.16$ & 
      $-19.02\pm0.16$ & ---\\
    $M_{V,\mathrm{cor}}$ & $-18.69\pm0.31$ & $-19.01\pm0.16$ & 
      $-19.04\pm0.16$ & ---\\ 
    $M_{I,\mathrm{cor}}$ & $-18.44\pm0.31$ & $-18.69\pm0.16$ & 
      $-18.74\pm0.16$ & ---\\
    \hline
  \end{tabular}
\end{table*}

A meaningful comparison of \object{SN\,1994D} with other SNe requires
the distances of the comparison objects to be derived with the same
method of GCLFs, because then the comparison is valid without the need
for an absolute calibration of GCLFs. Moreover, these SNe should be
well observed. These two conditions are fulfilled only in the Fornax
galaxy cluster with its 3 Ia SNe \object{SN\,1980N},
\object{SN\,1981D} (both in \object{NGC\,1316}), and
\object{SN\,1992A} (in \object{NGC\,1380}). The light curves of the
first two SNe have been presented by Hamuy et~al. (\cite{hamuy:91}),
while \object{SN\,1992A} (one of the best observed SNe) is described
by Suntzeff (\cite{suntzeff:96}).

All these SNe are relatively fast decliners, and thus follow the trend
already visible in the Cal\'an/Tololo SN sample (Hamuy et~al.
\cite{hamuy:96}) that spiral galaxies host all kinds of SNe, while
early-type galaxies tend to host only fast decliners (which are
intrinsically fainter).

The Fornax cluster is well suited to serve as a distance standard. It
is compact with hardly any substructure and consists in its core
region almost entirely of ellipticals and S0 galaxies, the
characteristics of a well relaxed galaxy cluster.  The distance based
on GCLFs has been determined through the work of Kohle et~al.
(\cite{kohle:96}) and Kissler-Patig et~al. (\cite{kissler:97a}), who
analysed the GCSs for several early-type galaxies. In addition,
Kissler-Patig et~al.  (\cite{kissler:97b}) and Della~Valle et~al.
(\cite{dellavalle:98}) determined the luminosity function for the GCS
of \object{NGC\,1380}, host of \object{SN\,1992A}.  Kohle et~al.
(\cite{kohle:96}) derived mean apparent TOMs for their sample of 5
Fornax galaxies of $V^0=23.67\pm0.06$.  This is in perfect agreement
with the TOM of \object{NGC\,1380} given by Della~Valle et~al.
(\cite{dellavalle:98}), which is $V^0=23.67\pm0.11$.  The difference
in distance moduli between Kohle et~al. (\cite{kohle:96}) and
Della~Valle et~al.  (\cite{dellavalle:98}) stems from a revised
absolute TOM of the Galactic GCS. If this is based on the Gratton
et~al.  (\cite{gratton:97}) relation, which is similar to our
eq.~(\ref{eq:mvrr}), the distance modulus is
$\mu_\mathrm{Fornax}=31.35\pm0.15$\,mag.  It is very satisfactory that
several other independent methods confirm our distance estimate
(Richtler et~al. \cite{richtler:99b}).  For example, Madore et~al.
(\cite{madore:98}) recently published a Cepheid distance to
\object{NGC\,1365} and quoted $31.35\pm0.2$\,mag.  Moreover, the
method of surface brightness fluctuations gives a Fornax distance
modulus of $31.32\pm0.24$\,mag (Jensen et~al.  \cite{jensen:98}).

For the other host galaxy in Fornax, \object{NGC\,1316}, no published
GCLF distance exists. Grillmair et~al.  (\cite{grillmair:99}), using
HST data, found no turn-over but an exponential shape of the GCLF.
They explained this with a population of open clusters mixed in,
presumably dating from the past merger event. Preliminary results from
ongoing work (G\'omez et~al., in preparation), however, place the TOM
of the entire system at the expected magnitude, but indicate a radial
dependence in the sense that the TOM becomes fainter at larger radii.
At this moment, we have no good explanation.  Perhaps, the different
findings correspond to the different regions in NGC\,1316
investigated. This issue is not yet settled.

An individual distance value for NGC\,1316 based on planetary nebulae
seems to indicate a shorter distance to this galaxy than to the Fornax
cluster.  McMillan et~al. (\cite{mcmillan:93}) find
$\mu_\mathrm{PN,NGC\,1316}=31.13$\,mag, but this value is based on a
M31 distance modulus of $24.26$\,mag.  Adopting the now widely
accepted M31 Cepheid distance modulus of $24.47$\,mag (e.g. Stanek \&
Garnavich \cite{stanek:98}, Kochanek \cite{kochanek:97}), the
NGC\,1316 modulus increases to $31.34$\,mag, perfectly consistent with
the global Fornax distance.  We therefore adopt the Fornax distance
modulus of $\mu_\mathrm{Fornax}=31.35\pm0.15$\,mag for
\object{SN\,1992A} \emph{and} \object{SN\,1980N}.

Table~\ref{tab:sndata} lists the absolute magnitudes of
\object{SN\,1994D} together with two Fornax cluster SNe
(\object{SN\,1981D} was badly observed), corrected for decline-rate
and colour. All values can be derived from the photometric data and
distances shown in table~\ref{tab:sndata} together with our correction
equation (\ref{eq:mcor}).

One notes that \object{SN\,1994D} seems to be 0.3\,mag too faint in
all three photometric bands. However, the error bars overlap with
those of the other SNe.  Thus, there is no reason to claim that this
SN is peculiarly dim.  On the other hand, if one takes Tonry's
unpublished SBF distance modulus, which is 0.28\,mag greater than
ours, all three SNe brightnesses would agree within $\pm0.05$\,mag.
At any rate, it is apparent that \object{SN\,1994D} is definitely not
overluminous and cannot be taken for a counter-example against the
general validity of the decline-rate--colour--luminosity relation.

\subsection{The Hubble constant}

Once the absolute brightness of a supernova is known, the Hubble
diagram of SNe Ia (Hamuy et~al. \cite{hamuy:96}) allows the derivation
of the Hubble constant with an accuracy which is largely determined
only by the error in the supernova brightness. The fact that the
Fornax SNe are well observed in combination with a reliable distance
makes these objects probably the best SNe for this purpose.  The
relation between the Hubble constant, the absolute corrected magnitude
$M$ and the zero-point $Z$ in (\ref{eq:mcor}) reads
\begin{equation}
  \label{eq:H0}
  \log \frac{H_0}{\mathrm{km\,s^{-1}\,Mpc^{-1}}}=0.2 \cdot(M-Z)+5.
\end{equation}

Taking the data from \object{SN\,1994D}, \object{SN\,1992A}, and
\object{SN\,1980N} presented in Table~\ref{tab:sndata} and the
zero-points from Table~\ref{tab:coeff} yields a Hubble constant of
\begin{equation}
  \label{eq:hc}
  H_0=75\pm6\pm6\,\mathrm{km\,s^{-1}\,Mpc^{-1}}.
\end{equation}
The systematic error originates from the horizontal branch magnitude,
the universality of the TOM, the adopted Fornax distance and the error
in $Z$. The statistical error is just the scatter from the three
corrected SNe absolute magnitudes. This Hubble constant value is
dominated by the data from the Fornax SNe because of the small
statistical weight of \object{SN\,1994D}.

\section{Summary}

We determined the GCLF distance of \object{NGC\,4526} to be
$12.0\pm3.9$\,Mpc, placing this galaxy in front of the Virgo cluster.
Although this distance is not precise enough to give a good absolute
magnitude it is sufficient to show that \object{SN\,1994D} is
\emph{not} an overluminous type Ia SN but rather a normally bright
event compared to the two Fornax SNe \object{SN\,1992A} and
\object{SN\,1980N}.

All GCLF distances depend on the absolute luminosities of the Galactic
GCs.  It is therefore vital to calibrate the absolute TOM properly to
minimise systematic errors. The most important calibration parameter
is the absolute magnitude of the horizontal branch which directly
enters our distance scale. The treatment of the Galactic GC sample
shows that the asymmetric distribution in $V$ and incompleteness
effects in the $I$ band can introduce a bias in the TOM of over
$0.1$\,mag.

With the distances and the apparent decline-rates and colour-corrected
magnitudes of the three SNe \object{SN\,1994D}, \object{SN\,1982A}, and
\object{SN\,1980N} one obtains an absolute corrected peak magnitude of
type Ia SNe. The Cal\'an/Tololo SN sample gives us the zero-point of
the Hubble diagram, the second ingredient in calculating the Hubble
constant from type Ia SNe. Our data give
$H_0=75\pm6(\mbox{sys.})\pm6(\mbox{stat.})\,\mathrm{km\,s^{-1}\,Mpc^{-1}}$.

\begin{acknowledgements}
  We thank M.~Kissler-Patig, T.~Puzia, M.~G\'omez, and
  W.~\mbox{Seggewiss} for useful discussions and G.~Ogilvie and
  C.~Kaiser for reading the manuscript. The referee
  Dr.~W.~\mbox{Richmond} helped considerably to improve the paper's
  clarity. We also thank M.~Della~Valle for pointing out to us a
  mistake in an earlier version.
  This paper uses the MDS: The Medium Deep Survey catalogue is based
  on observations with the NASA/ESA Hubble Space Telescope, obtained
  at the Space Telescope Science Institute, which is operated by the
  Association of Universities for Research in Astronomy, Inc., under
  NASA contract NAS5-26555. The Medium-Deep Survey analysis was funded
  by the HST WFPC2 Team and STScI grants GO2684, GO6951, and GO7536 to
  Prof.~Richard Griffiths and Dr.~Kavan Ratnatunga at Carnegie Mellon
  University.
\end{acknowledgements}

\end{document}